# Noncovalent Interactions in Supramolecular Complexes: A Study on Corannulene and the Double Concave Buckycatcher

*Bryan M. Wong**

Materials Chemistry Department, Sandia National Laboratories, Livermore, California 94551

*Corresponding author. E-mail: bmwong@sandia.gov.

Stimulated by the recent observation of π-π interactions between $C_{60}$ and corannulene subunits in a molecular tweezer arrangement (*J. Am. Chem. Soc.* **2007**, *129*, 3842), a density functional theory study was performed to analyze the electronic structure and properties of various noncovalent corannulene complexes. The theoretical approach is first applied to corannulene complexes with a series of benchmark molecules ($CH_4$, $NH_3$, and $H_2O$) using several new-generation density functionals. The performance of nine density functionals, illustrated by computing binding energies of the corannulene complexes, demonstrates that Zhao and Truhlar's MPWB1K and M05-2X functionals provide energies similar to that obtained at the SCS-MP2 level. In contrast, most of the other popular density functionals fail to describe this noncovalent interaction or yield purely repulsive interactions. Further investigations with the M05-2X functional show that the binding energy of $C_{60}$ with corannulene subunits in the relaxed molecular receptor clip geometry is -20.67 kcal/mol. The results of this calculation further support the experimental interpretation of pure π-π interactions between a convex fullerene and the concave surfaces of two corannulene subunits.

**I. Introduction**

Noncovalent interactions are ubiquitous in chemistry and are a main source of stability for many molecular complexes in nanoscience, materials chemistry, and biochemistry[1-3]. An area where noncovalent interactions are particularly important is the self-assembly of carbon nanostructures in which van der Waals effects promote the formation of multiwall carbon nanotubes and multishell fullerenes[4]. One of the most interesting applications of these supramolecular complexes is their potential to act as chemical receptors for molecular recognition[5,6]. For example, the π electron system in metallic carbon nanotubes demonstrates a high selectivity and affinity to specific neutral and charge-transfer aromatic molecules[7]. Moreover, the adsorbed molecules can be easily removed since the molecule-nanotube interactions are noncovalent. These highly specific yet reversible interactions suggest a first step towards nanosensor devices which are potentially recyclable.

A particularly interesting feature of interacting π systems is the favorable attraction between concave polycyclic aromatic hydrocarbons with convex fullerene cages. The interactions in these concave-convex complexes are unusual because the π orbitals between the curved surfaces are highly polarized, unlike the nearly uniform density of π electrons in graphene sheets[8]. As a result, concave carbon surfaces have been calculated to be even better π electron donors than their planar counterparts[9,10]. Recognizing this favorable concave-convex interaction, Sygula et al. recently isolated an unconventional complex of $C_{60}$ with $C_{60}H_{28}$, a "molecular tweezer" with two corannulene subunits[11] depicted in Figure 1. Using NMR titration and X-ray structure techniques, Sygula and co-workers analyzed crystal structures for the complex and provided experimental confirmation of pure concave-convex π-π interactions.

Despite the numerous experimental studies on noncovalently interacting systems, the progress of *ab initio* calculations has been slower in quantitatively describing their unique electronic structure. The difficulties stem from the intricate evaluation of dispersion forces which arise from instantaneous multipole/induced multipole charge fluctuations between molecular surfaces[12]. Since the Hartree-Fock method is a mean-field theory which only describes average electronic effects, it is incapable of capturing the instantaneous electron interactions which give rise to dispersion forces[13]. The conventional approach for handling these systems is to use Møller-Plesset perturbative theory (MP2) in conjunction with highly correlated coupled-cluster methods (CCSD(T)) to estimate their binding energies[14]. However, the computational cost of CCSD(T) methods is too high for routine application to molecules larger than about 20 atoms.

Density functional theory (DFT), on the other hand, scales efficiently with molecular size and is the method of choice for studying large molecular systems. Unfortunately, the accurate description of dispersion forces has been a traditional failure of most current density functionals, and the development of new methods for properly treating noncovalent interactions is still a topic of active research[15-19]. Responding to this need for better-performing functionals, Zhao and Truhlar recently proposed a hybrid meta-GGA functional, M05-2X, for computing $\pi$-$\pi$ stacking and alkane isomerization energies[20]. The M05-2X functional remedies the deficiencies of other hybrid functionals by incorporating an improved treatment of spin kinetic energy density in both the exchange and correlation functionals. This simultaneous optimization of exchange and correlation functionals accounts for medium-range correlation energy which is a source of error in van der Waals systems. Based on extensive tests consisting of several small van der Waals dimers[21-23], the authors further concluded that M05-2X is the best functional for predicting geometries and energies of noncovalently bound systems. Wodrich et al. have independently compared M05-2X bond separation energies against experimental data for 72 hydrocarbons and found that M05-2X has excellent across-the-board performance for optimized energies and geometries[24]. Recently, in a time dependent density functional study, Santoro and co-workers further demonstrated that the M05-2X functional also provides an accurate description of excited states in $\pi$-stacked nucleobases[25].

Since noncovalent interactions between curved aromatic hydrocarbons and fullerenes can serve as models for self-assembling nanostructures, there is a need for reliable methods to estimate the binding energies of noncovalently bound systems. Furthermore, since it is difficult to extract the binding energy of stacked complexes from experiment, it is useful to compare recently developed DFT methods with results obtained from post-Hartree-Fock wavefunction-based methods. In this work, the structures and electronic properties of various benchmark corannulene complexes (corannulene attached to methane, ammonia, and water) are computed using nine density functionals. Mean signed deviations (MSDs) of binding energies from other wavefunction-based methods are used to further evaluate the performance of these functionals. These calculations supplement existing benchmark calculations of noncovalent interactions and provide guidance on computing the binding energy of the $C_{60}\cdots C_{60}H_{28}$ buckycatcher complex. In addition, other electronic properties such as natural bond orbital charges are also examined to assess the stability of π-π interactions in the buckycatcher complex.

## II. Computational Details

All binding energies of corannulene with $CH_4$, $NH_3$, and $H_2O$, were computed using the 6-311+G(d,p) basis set which has previously shown sufficient accuracy for other corannulene complexes[26]. Calculations on the $C_{60}\cdots C_{60}H_{28}$ buckycatcher complex were performed with an augmented 6-311G(d,p) basis set which has an extra diffuse function only on the corannulene subunits (five diffuse exponents were symmetrically placed in the interior of each corannulene subunit). Geometry optimizations for all molecules were unconstrained, allowing for deformation and optimal intermolecular distances in the fully relaxed complex. The convergence criteria for maximum and root-mean-square forces were set to $4.5 \times 10^{-4}$ Hartree/Bohr and $3.0 \times 10^{-4}$ Hartree/Bohr respectively. In order to correct for the basis set incompleteness which arises from using finite atom-centered basis sets on each monomer, the counterpoise correction[27] was applied to all reported binding energies.

The nine functionals utilized in the present analysis include the most widely used hybrid functional, B3LYP[28], the parameter-free PBE[29] and PW91[29] functionals, B3PW91[28,30], B1B95[31], MPW1PW91[32], and Truhlar's recent meta-GGA MPW1B95[33], MPWB1K[33], and M05-2X[20] functionals. All *ab initio* calculations were performed at the National Center for Supercomputing Applications with the NWChem 5.0 software package developed by Pacific Northwest National Laboratories[34].

**III. Results and Discussion**

**3.1. Corannulene complexes with $CH_4$, $NH_3$, and $H_2O$.** Since the interpretation of DFT results on large molecular systems is never straightforward, it is important to compare current functionals against results obtained from using sophisticated post-Hartree Fock calculations. It should be mentioned that Zhao and Truhlar have already shown that the M05-2X functional accurately describes van der Waals complexes and $\pi$-stacking interactions in a large set of benchmark calculations[20-23]. To supplement their extensive study, addition calculations are presented on medium-sized complexes involving noncovalent interactions in corannulene complexes. Fortunately, the very recent publication of a detailed spin-component-scaled MP2 (SCS-MP2) study on corannulene complexes provides a good benchmark comparison with DFT results[35]. Grimme and co-workers have successfully applied the SCS-MP2 method to several intermolecular interactions involving $\pi$-stacking[36-39], and their data on corannulene complexes is used here as reference values.

Figures 2 (a)-(c) show the optimized geometries of the benchmark corannulene complexes with $CH_4$, $NH_3$, and $H_2O$ considered in the present work. As can be seen in Figure 2 (d), the electrostatic potential is more negative in the center of corannulene than on the outside rim. Consequently, geometry optimizations for all the molecules place their hydrogen atoms, rather than the hetero atoms, pointing towards the corannulene. The counterpoise-corrected binding energies obtained from the various density functionals are reported with the corresponding SCS-MP2 energies in Table 1.

The expected trend for all DFT methods is that the polar $NH_3$ and $H_2O$ molecules bind more strongly to corannulene than the nonpolar $CH_4$. Despite this common prediction, the performances of each of the nine DFT methods relative to each other are quite different. Both B3LYP and B3PW91, which are based on Becke's three parameter exchange functional, yield repulsive interactions for the $CH_4$···corannulene complex. In general, B3LYP, B3PW91, and MPW1PW91 predict very weak complexation energies for all three molecules. The mean signed deviations (MSD) from SCS-MP2 values indicate that PW91, PBE, B1B95, and MPW1B95 give slightly better performance but still yield fairly large errors. The M05-2X functional, closely followed by MPWB1K, has the lowest MSD and gives an average error of about 16%. All other DFT methods, including the most widely used B3LYP functional, produce much larger deviations. In addition, the B3LYP functional is known to produce large errors with increased system size[24], but the results in this part of the study demonstrate that the M05-2X functional still provides reliable results for fairly large systems.

**3.2. $C_{60}\cdots C_{60}H_{28}$ buckycatcher complex.** For the initial studies on the buckycatcher complex, a $C_{60}$ molecule was placed between the two corannulene subunits, and an unconstrained geometry optimization was performed. Since analytical and numerical frequencies are not currently implemented in NWChem for many of the newer density functionals, the buckycatcher complex was first optimized at the PBE/6-31G level of theory. Harmonic frequency calculations were performed at the same level of theory on the equilibrium structure, characterizing it as a minimum on the potential energy surface. The PBE/6-31G equilibrium structure was then used as the starting initial geometry for unconstrained 6-311G(d,p) (augmented with ten corannulene-centered diffuse functions) optimizations using the other nine DFT methods. The counterpoise-corrected binding energies, $\Delta E$, and geometries obtained from the various density functionals are reported in Table 2.

Using the initial PBE/6-31G relaxed structure, all DFT optimizations converged to a $C_{2v}$ structure where the five- and six-membered rings in the corannulene subunits and $C_{60}$ are nearly eclipsed (Figures 1 and 3). The equilibrium $C_{60}\cdots$corannulene subunit distance (defined as the shortest atom-to-atom distance) for each of the nine DFT methods is listed in Table 2. These optimized geometric parameters can be directly compared with the shortest atom-to-atom distance of 3.128 Å obtained from the experimental X-ray crystal structure. The M05-2X calculated distance of 3.20 Å is in excellent agreement with experiment and gives the least error (around 2.3%) compared to all the other functionals. MPWB1K, MPW1B95, and B1B95 also perform satisfactorily with geometric errors of less than 7%. However, the B3LYP and B3PW91 functionals give poor results and compute the largest deviations for the buckycatcher complex (errors of 104% and 129% respectively).

A comparison of binding energies parallels the trend reported for the optimized geometries. As expected from the previous analysis of corannulene complexes with $CH_4$, both B3LYP and B3PW91 yield positive binding energies and also fail to describe the π-π stacking in the buckycatcher complex. The use of the MPW1PW91 and PBE functionals also results in large repulsive interactions. Consistent with the previous results on corannulene, the B1B95 and PW91 functionals give stronger binding energies, but the interactions are still weakly bound. Finally, M05-2X and MPWB1K produce the strongest binding energies (most negative) with the M05-2X value almost twice that obtained with the MPWB1K functional. The large binding energy of -20.67 kcal mol$^{-1}$ compares very well with a current study on π-stacking interactions involving fullerenes. For example, Zhao and Truhlar recently demonstrated that the binding energy of $C_{60}$ inside a hydrocarbon nanoring is -28.0 kcal mol$^{-1,40}$. The binding energy of the $C_{60}\cdots C_{60}H_{28}$ buckycatcher complex is 74% of this value. This is reasonable since the nanoring completely encapsulates $C_{60}$ whereas only about half of the $C_{60}$ surface is noncovalently bound in the buckycatcher complex.

In order to further explain these trends, a natural bond orbital (NBO) analysis[41] was performed for all nine DFT methods. The NBO procedure uses only the information from the density matrix of the wavefunction and yields a set of localized orbitals which give the most accurate Lewis-like description of the total electron density. Moreover, the NBO method is not prone to basis set errors such as those produced by Mulliken population analyses which often behave erratically with large basis sets[42]. For all the DFT methods exhibiting large negative binding energies, the NBO analysis points strongly to π → π* interactions from $C_{60}H_{28}$ to $C_{60}$. This π → π* interaction involves weak delocalizations from π bonds on the corannulene subunits into near unfilled π* antibonding orbitals of the $C_{60}$ molecule. These occupancy shifts from filled orbitals of one molecule to the unfilled orbitals of the other are a hallmark of charge transfer interactions (Figure 3). Indeed, a natural population analysis of the charge transferred between $C_{60}$ and $C_{60}H_{28}$ reported in Table 2 indicates that large values of transferred charge are associated with significant energy stability. A rough rule of thumb often used is that the natural bond orbital charge is directly proportional to the energy stabilization associated with the charge transfer[41]. Consequently, $\Delta E$ is considerably greater in magnitude for M05-2X and MPWB1K, while in the non-binding B3LYP and B3PW91 complexes, both $\Delta E$ and $q$ are significantly reduced. In general, the trends predicted by Truhlar's M05-2X functional are consistent with experimental structures and high-level calculations. These trends also give quantitative evidence of convex surfaces preferring to act as $\pi$-electron acceptors and concave surfaces acting as $\pi$-electron donors.

## IV. Conclusion

Systematic *ab initio* investigations were performed for several corannulene complexes starting with $CH_4$, $NH_3$, and $H_2O$ up to a complex with $C_{60}$ in a molecular tweezer geometry. The molecular structures, binding energies, and electrostatic properties were obtained using a set of DFT methods which include new-generation functionals with improved performance for noncovalent interactions. In agreement with previous benchmark calculations, the current study indicates that Truhlar's MPWB1K and M05-2X functionals are the only viable DFT methods for accurately describing noncovalent binding energies in corannulene systems. All other functionals exhibit a tendency to underestimate binding energies due to their poor description of dispersion interactions.

Among the noncovalent interactions studied, the buckycatcher supramolecular complex presents an interesting yet difficult test for density functional theory. Using the same set of DFT methods, the fully optimized structure and electrostatic properties of the buckycatcher complex were obtained using quantum chemical methods for the first time. A comparison with an X-ray crystal structure demonstrates that the M05-2X functional reproduces the experimental geometry very well and predicts a binding energy of -20.67 kcal/mol with an augmented 6-311G(d,p) basis. In addition, a natural bond orbital analysis was performed to understand the stability of the buckycatcher complex based on the strength of charge transfer interactions. The NBO charge transfer analysis and electrostatic potential plots are consistent with previous experimental observations on $C_{60}$ and corannulene. That is, concave $\pi$ surfaces are electron-rich while the convex surfaces of $C_{60}$ are naturally electron-deficient, an experimental observation demonstrated by the ease with which $C_{60}$ forms complexes with various electron-rich metals. The results of this *ab initio* characterization further support the experimental interpretation of pure π-π interactions between a convex fullerene and the concave surfaces of two corannulene subunits.

In summary, the results provide indications that supramolecular behavior in molecular nanostructures can be described with good accuracy when using a suitable DFT method. The new-generation density functionals used in this work show great promise for studying large systems since they give reasonable results at a much lower computational cost than other wavefunction-based methods. However, widely used density functionals like B3LYP can not describe the long-range electron correlations which are responsible for dispersion forces. In view of the huge popularity of B3LYP (which has over 20,000 citations to date), existing calculations of weakly-bound systems using this particular functional may need to be re-examined. As concerns the choice of the DFT method, Truhlar's M05-2X functional is recommended for describing noncovalent interactions in molecular nanostructures.

**Acknowledgement.** This work was partially supported by the National Center for Supercomputing Applications under grant number TG-CHE070084N and utilized the NCSA Cobalt SGI Altix System. Sandia is a multiprogram laboratory operated by Sandia Corporation, a Lockheed Martin Company, for the United States Department of Energy's National Nuclear Security Administration under contract DE-AC04-94AL85000.

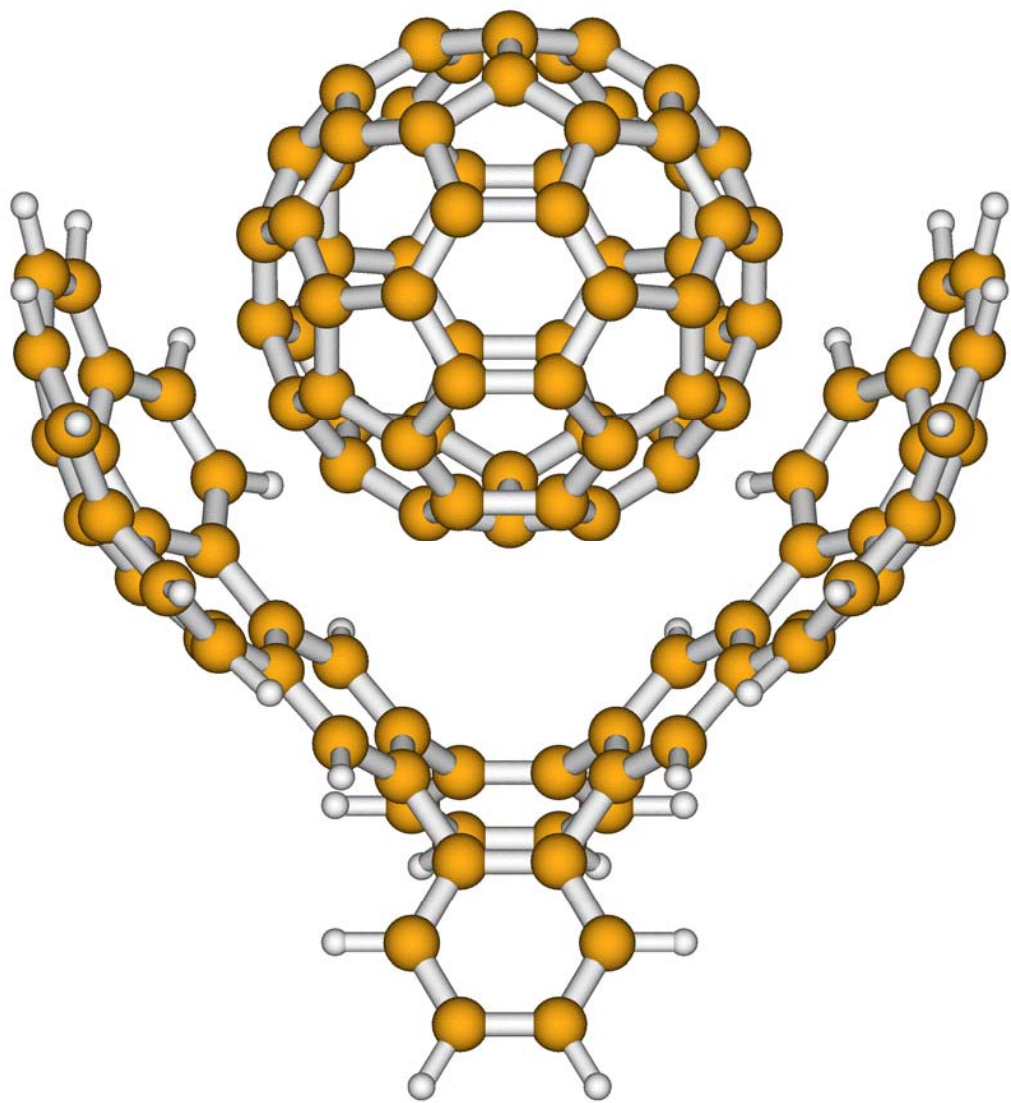

**Figure 1.** Optimized geometry of the $C_{60}\cdots C_{60}H_{28}$ buckycatcher complex at the M05-2X level of theory with an augmented 6-311G(d,p) basis set. The corannulene subunits are in van der Waals contact with the fullerene molecule at a distance of 3.20 Å (shortest atom-to-atom distance).

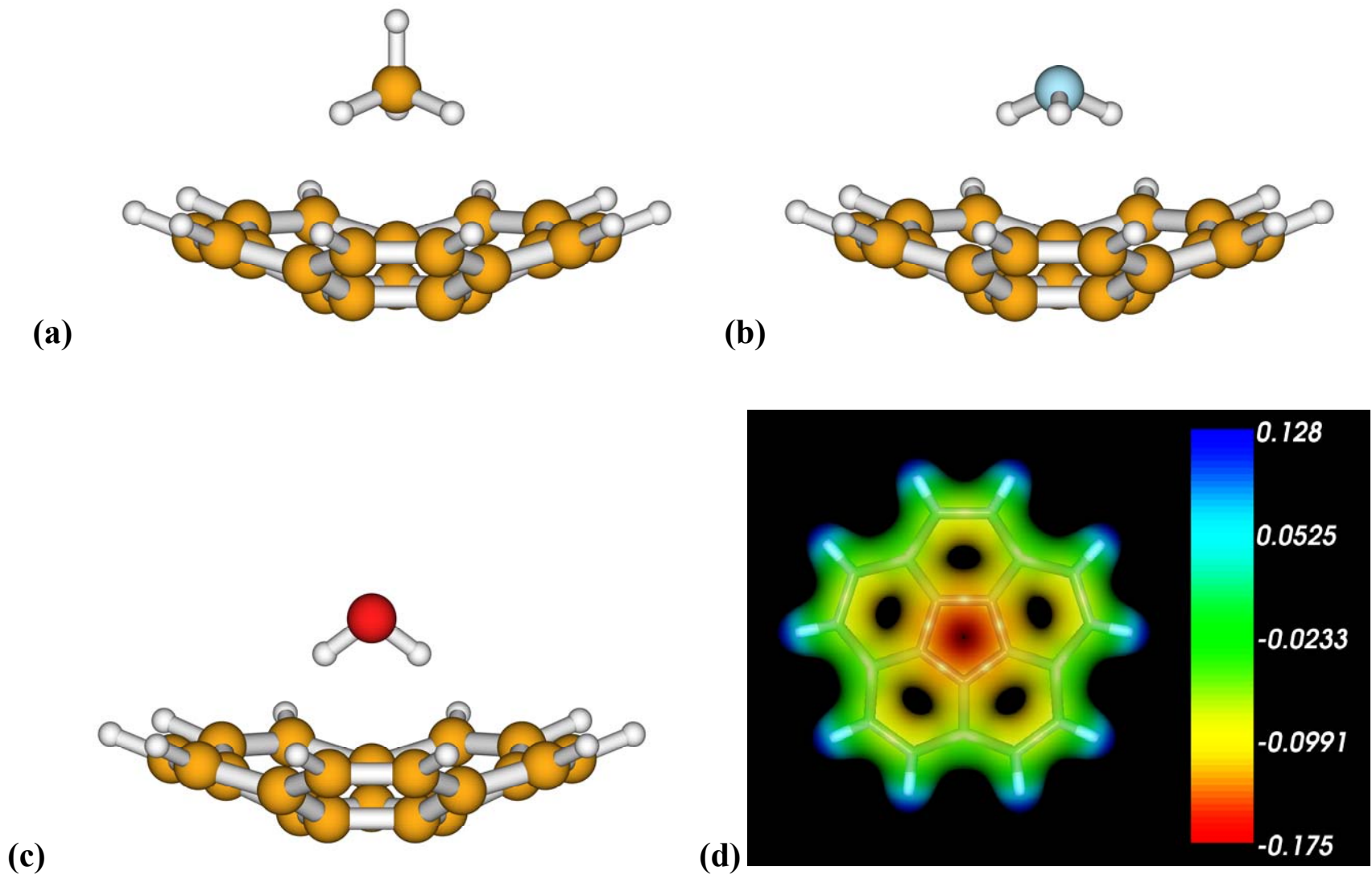

**Figure 2.** Optimized geometries of corannulene complexes with (a) methane ($CH_4$), (b) ammonia ($NH_3$), and (c) water ($H_2O$) at the M05-2X/6-311G+(d,p) level of theory. In Figure 2 (d), the electrostatic potential of an isolated corannulene molecule is shown. The center of the corannulene bowl has a lower electrostatic potential than the outside rim.

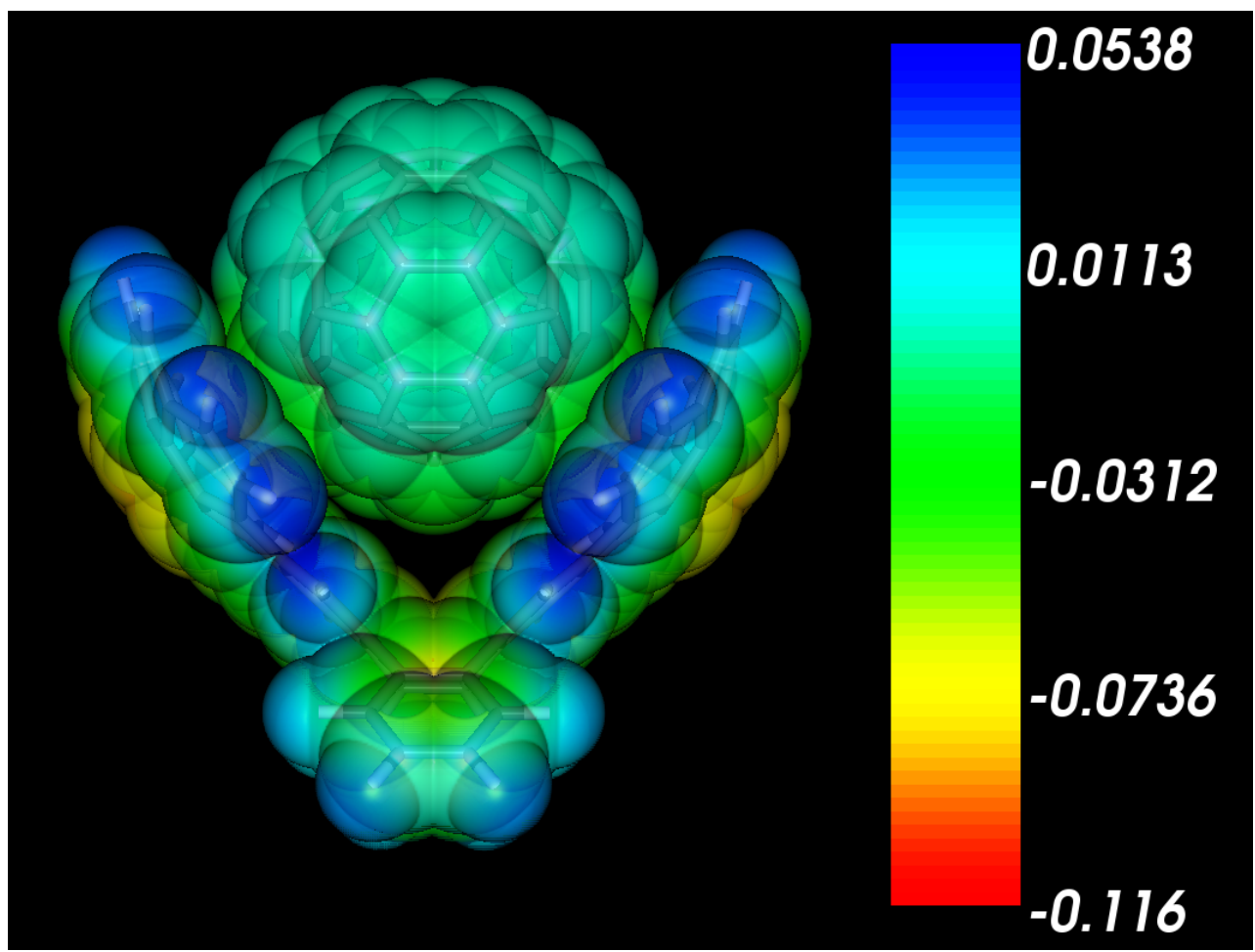

**Figure 3.** Electrostatic potential of the $C_{60}\cdots C_{60}H_{28}$ buckycatcher complex at the M05-2X/6-311G(d,p) level of theory with an augmented 6-311G(d,p) basis set. Since the concave $\pi$ surfaces of corannulene are electron-rich while the convex surfaces of $C_{60}$ are naturally electron-deficient, regions of the fullerene cage near the corannulene subunits are more negatively charged than the other free surfaces.

TABLE 1: Binding Energies (in kcal mol$^{-1}$) for Corranulene Complexes with $CH_4$, $NH_3$, and $H_2O$ Obtained with Various Density Functionals[a]

|  | B1B95 | B3LYP | B3PW91 | M05-2X | MPW1B95 | MPW1PW91 | MPWB1K | PBE | PW91 | SCS-MP2[b] |
|---|---|---|---|---|---|---|---|---|---|---|
| $CH_4$ | -1.18 | 0.04 | 0.06 | -4.20 | -2.64 | -0.12 | -2.95 | -0.48 | -0.91 | -4.3 |
| $NH_3$ | -2.51 | -0.44 | -0.14 | -5.89 | -3.86 | -0.91 | -4.26 | -1.72 | -2.10 | -4.8 |
| $H_2O$ | -2.32 | -0.93 | -0.71 | -5.88 | -3.59 | -1.43 | -3.96 | -2.15 | -2.50 | -4.6 |
| MSD | 2.56 | 4.12 | 4.30 | -0.75 | 1.20 | 3.75 | 0.84 | 3.12 | 2.73 | — |

[a] All DFT energies are counterpoise-corrected using 6-311G+(d,p) optimized geometries. [b] Best estimate from Ref. 34. [c] Abbreviation: Mean signed deviation.

**TABLE 2: Binding Energy ($\Delta E$), Equilibrium $C_{60}\cdots C_{60}H_{28}$ Distance ($d$), and Natural Bond Orbital Charge of $C_{60}$ ($q$) for the Buckycatcher Complex Obtained with Various Density Functionals**

|  | B1B95 | B3LYP | B3PW91 | M05-2X | MPW1B95 | MPW1PW91 | MPWB1K | PBE | PW91 |
|---|---|---|---|---|---|---|---|---|---|
| $\Delta E$ (kcal mol$^{-1}$)[a] | -1.88 | 0.34 | 0.09 | -20.67 | -7.74 | 0.75 | -10.22 | 0.91 | -0.45 |
| $d$ (Å)[b] | 3.34 | 6.38 | 7.16 | 3.20 | 3.32 | 4.50 | 3.31 | 3.72 | 3.69 |
| % deviation[c] | 6.8 | 104.0 | 128.9 | 2.3 | 6.1 | 43.9 | 5.8 | 18.9 | 18.0 |
| $q$ (electrons) | -0.0259 | 0.0009 | 0.0004 | -0.0389 | -0.0260 | -0.0021 | -0.0238 | -0.0078 | -0.0083 |

[a] All energies are counterpoise-corrected using 6-311G(d,p) optimized geometries. [b] Defined as the shortest atom-to-atom distance between $C_{60}$ and $C_{60}H_{28}$. [c] Percent deviation from X-ray data taken from Ref. 11.

(1) Claessens, C. G.; Stoddart, J. F. J Phys Org Chem 1997, 10, 254.

(2) Fyfe, M. C. T.; Stoddart, J. F. Acc Chem Res 1997, 10, 393.

(3) Kannan, N.; Vishveshwara, S. Protein Eng 2000, 13, 753.

(4) Mordkovich, V. Z. Chem Mater 2000, 12, 2813.

(5) Klärner, F.-G.; Kahlert, B. Acc Chem Res 2003, 36, 919.

(6) Harmata, M. Acc Chem Res 2004, 37, 862.

(7) Lu, J.; Nagase, S.; Zhang, X.; Wang, D.; Ni, M.; Maeda, Y.; Wakahara, T.; Nakahodo, T.; Tsuchiya, T.; Akasaka, T.; Gao, Z.; Yu, D.; Ye, H.; Mei, W. N.; Zhou, Y. J Am Chem Soc 2006, 128, 5114.

(8) Kawase, T.; Kurata, H. Chem Rev 2006, 106, 5250.

(9) Klärner, F.-G.; Panitzky, J.; Preda, D.; Scott, L. T. J Mol Mod 2000, 6, 318.

(10) Ansems, R. B. M.; Scott, L. T. J Phys Org Chem 2004, 17, 819.

(11) Sygula, A.; Fronczek, F. R.; Sygula, R.; Rabideau, P. W.; Olmstead, M. M. J Am Chem Soc 2007, 129, 3842.

(12) Tsuzuki, S.; Uchimaru, T.; Matsumura, K.; Mikami, M.; Tanabe, K. Chem Phys Lett 2000, 319, 547.

(13) Tsuzuki, S.; Lüthi, H. P. J Chem Phys 2001, 114, 3949.

(14) Sinnokrot, M. O.; Valeev, E. F.; Sherrill, C. D. J Am Chem Soc 2002, 124, 10887.

(15) Grimme, S.; J Comput Chem, 2004, 25, 1463.

(16) Grimme, S.; J Comput Chem, 2006, 27, 1787.

(17) Grimme, S.; Mück-Lichtenfeld, C.; Antony, J. J Phys Chem C 2007, 111, 11199.

(18) Staroverov, V. N.; Scuseria, G. E.; Tao, J.; Perdew, J. P. J Chem Phys 2003, 119, 12129.

(19) Tao, J.; Perdew, J. P. J Chem Phys 2005, 122, 114102-1.

(20) Zhao, Y.; Schultz, N. E.; Truhlar, D. G. J Chem Theory Comput 2006, 2, 364.

(21) Zhao, Y.; Truhlar, D. G. J Phys Chem A 2005, 109, 4209.

(22) Zhao, Y.; Truhlar, D. G. Phys Chem Chem Phys 2005, 7, 2701.

(23) Zhao, Y.; Truhlar, D. G. J Phys Chem A 2006, 110, 5121.

(24) Wodrich, M. D.; Corminboeuf, C.; Schreiner, P. R.; Fokin, A. A.; Schleyer, P. v. R. Org Lett 2007, 9, 1851.

(25) Santoro, F.; Barone, V.; Improta, R. J Comput Chem 2008, in press.

(26) Kandalam, A. K.; Rao, B. K.; Jena, P. J Phys Chem A 2005, 109, 9220.

(27) Boys, S. F.; Bernardi, F. Mol Phys 1970, 19, 553.

(28) Becke, A. D. J Chem Phys 1993, 98, 5648.

(29) Perdew, J. P.; Burke, K.; Ernzerhof, M. Phys Rev Lett 1996, 77, 3865.

(30) Perdew, J. P. Electronic Structure of Solids '91; Akademie Verlag: Berlin, 1991.

(31) Becke, A. D. Phys Rev A 1988, 38, 3098.

(32) Adamo, C.; Barone, V. J Chem Phys 1998, 108, 664.

(33) Zhao, Y.; Truhlar, D. G. J Phys Chem A 2004, 108, 6908.

(34) Bylaska, E. J.; de Jong, W. A.; Kowalski, K.; Straatsma, T. P.; Valiev, M.; Wang, D.; Aprà, E.; Windus, T. L.; Hirata, S.; Hackler, M. T.; Zhao, Y.; Fan, P.-D.; Harrison, R. J.; Dupuis, M.; Smith, D. M. A.; Nieplocha, J.; Tipparaju, V.; Krishnan, M.; Auer, A. A.; Nooijen, M.; Brown, E.; Cisneros, G.;

Fann, G. I.; Früchtl, H.; Garza, J.; Hirao, K.; Kendall, R.; Nichols, J. A.; Tsemekhman, K.; Wolinski, K.; Anchell, J.; Bernholdt, D.; Borowski, P.; Clark, T.; Clerc, D.; Dachsel, H.; Deegan, M.; Dyall, K.; Elwood, D.; Glendening, E.; Gutowski, M.; Hess, A.; Jaffe, J.; Johnson, B.; Ju, J.; Kobayashi, R.; Kutteh, R.; Lin, Z.; Littlefield, R.; Long, X.; Meng, B.; Nakajima, T.; Niu, S.; Pollack, L.; Rosing, M.; Sandrone, G.; Stave, M.; Taylor, H.; Thomas, G.; van Lenthe, J.; Wong, A.; Zhang, Z. "NWChem, A Computational Chemistry Package for Parallel Computers, Version 5.0" (2006), Pacific Northwest National Laboratory, Richland, Washington 99352-0999, USA.

(35) Grimme, S.; Antony, J.; Schwabe, T.; Müch-Lichtenfeld, C. Org Biomol Chem 2007, 5, 741.

(36) Grimme, S. J Chem Phys 2003, 118, 9095.

(37) Grimme, S. Chem Eur J 2004, 10, 3423.

(38) Piacenza, M.; Grimme, S. Chem Phys Chem 2005, 6, 1554.

(39) Piacenza, M.; Grimme, S. J Am Chem Soc 2005, 127, 14841.

(40) Zhao, Y.; Truhlar, D. G. J Am Chem Soc 2007, 129, 8440.

(41) Foster, J. P.; Weinhold, F. J Am Chem Soc 1980, 102, 7211.

(42) Reed, A. E.; Weinhold F.; Curtiss, L. A.; Pochatko, D. J. J Chem Phys 1986, 84, 5687.